# Identification of two trapping mechanisms responsible of the threshold voltage variation in SiO$_2$/4H-SiC MOSFETs

Patrick Fiorenza[1] [*], Filippo Giannazzo[1], Mario Saggio[2], Fabrizio Roccaforte[1]

1) Consiglio Nazionale delle Ricerche – Istituto per la Microelettronica e Microsistemi (CNR-IMM), Strada VIII, n. 5 - Zona Industriale, 95121 Catania, Italy

2) STMicroelectronics, Stradale Primosole n. 50 - Zona Industriale, 95121 Catania, Italy

[*]E-mail: patrick.fiorenza@imm.cnr.it

## Abstract

A non-relaxing method based on cyclic gate bias stress is used to probe the interface or near-interface traps in the SiO$_2$/4H-SiC system over the whole 4H-SiC band gap. The temperature dependent instability of the threshold voltage in lateral MOSFETs is investigated and two separated trapping mechanisms were found. One mechanism is nearly temperature independent and it is correlated to the presence of near interface oxide traps that are trapped via tunneling from the semiconductor. The second mechanism, having an activation energy of 0.1 eV, has been correlated to the presence of intrinsic defects at the SiO$_2$/4H-SiC interface.

4H-SiC devices technology is still affected by some reliability concerns [1]. As an example, the SiC community is focused on the comprehension of the threshold voltage ($V_{th}$) instability phenomena, often observed in 4H-SiC MOSFETs [2,3]. Such $V_{th}$ instability effects are likely due to electrons tunneling into and out of near-interfacial oxide traps (NIOTs) that extend spatially into the gate oxide from the SiC interface [4,5]. Usually, the $SiO_2$/4H-SiC interface physics is studied on simple MOS capacitors [6,7,8], but the full picture on the real impact of the trapping states can be achieved only studying the real MOSFET device [9].

In this context, we have recently studied the discharge of the NIOTs in later MOSFETs by transient gate-capacitance [5] and gate-current [10,11] measurements. However, the separation of each trapping contribution is fundamental for the comprehension of the $SiO_2$/4H-SiC interface physics. In particular, non-relaxing methods are needed in order to minimize the amount of traps undetected during the investigation. Hence, the role of the $SiO_2$/4H-SiC interface physics is the key factor to control undesired effects of the $V_{th}$ instability in power MOSFETs. In particular, *Sometani et al.* [12] have presented a non-relaxation method that allows to measure the $V_{th}$ using a fixed $V_G$ and adjusting the drain-source bias ($V_{DS}$) value keeping constant the $I_D$. Hence, they have found that the trapped charges with a small detrapping time constant can only be observed in the non-relaxation method. In this context, it results clear that methods that minimize the trap relaxation are needed to access at the intimate mechanisms involved in the threshold voltage instability, i.e. reducing the time between the stress and the measurements and avoiding voltage swings that can hide the stress effect.

Firstly, it is important to understand how to define the threshold voltage ($V_{th}$). The standard procedure to determine $V_{th}$ is the linear fit of the $I_D^{0.5}$ vs $V_G$. However, this method requires the sweep of the gate bias $V_G$ and is essentially insensitive to charge state variation in the gate insulator during the measurement. Since the oxide traps that are very close to the interface will change their charge state not only during the stress but also during the measurement, it is important to monitor the $V_{th}$ variation keeping constant the applied gate bias in order to quantify also the amount of charge trapped in the oxide during the device qualification process. Hence, it is important to minimize the system perturbation due to the $V_{th}$ measurement.

In this work, we propose a non-relaxing method based on cyclic gate bias stress that allows to probe the whole 4H-SiC band gap and to separate the contribution of different traps avoiding voltage swings and reducing the time between the stress and the measurement itself. In particular, the drain current at a single gate bias value ($I_D$-$V_{G\text{-}read}$) measurements were used to estimate the variation of the $V_{th}$ in 4H-SiC lateral MOSFETs after both positive and negative gate bias stress. The interpretation of the experimental results allowed to separate the interface states ($N_{it}$) and the near interface oxide traps (NIOTs) to the $V_{th}$ variation.

Lateral MOSFETs were fabricated on 4°-off-axis n-type (0001) 4H-SiC epitaxial layers ($1\times10^{16}$ cm$^{-3}$) and an Al-implanted body region ($N_A\sim10^{17}$cm$^{-3}$). The gate oxide was a 40nm thick SiO$_2$ layer [13]. The current voltage ($I_D$-$V_G$) and the drain current transient ($I_D$-t) characteristics of the devices were measured in a CASCADE Microtech probe station, using a Keysight B1505A parameter analyser.

Fig. 1 shows the $I_D^{0.5}$-$V_G$ curves collected in a lateral MOSFET at a fixed drain voltage by increasing the gate bias ($V_G$) up to +30V. The saturation drain current can be described by the following equation [14]:

$$(I_D^{sat})^{0.5} = \sqrt{\frac{mW}{L}\mu C_{ox}}(V_G - V_{th}) \equiv \sqrt{\alpha_{sat}} \times (V_G - V_{th}) \qquad (1)$$

where *m* is a function of doping density; *W* is the channel width, *L* is the channel length, $C_{ox}$ is the gate capacitance and *μ* is the electron mobility in the channel [14]. It is clear that for large $V_D$ values, Eq. 1 can be used to fit the experimental data, and the $V_{th}$ value can be determined as the intercept with the gate bias axis.

In the presence of $V_{th}$ instabilities, it is possible to monitor the time dependence of the saturation current $I_D^{sat}(t)$. This transient current variation implies that also a threshold voltage variation occurred:

$$V_{th}(t) = V_G - \sqrt{\frac{I_D^{sat}(t)}{\alpha_{sat}}} \qquad (2).$$

The insert in Fig. 1 shows the experimental transient drain current ($I_D$-t) collected at room temperature at fixed $V_D$=+5 V and $V_G$=+20 V on a fresh MOSFET. As can be seen the drain current decreases with the time, thus hinting at an effective variation of the $V_{th}$ during the measurement.

Fig. 2 helps to visualize how the variation of the $I_D$ can be correlated to the $V_{th}$ variation. In fact, assuming a rigid shift of the transfer characteristic of the MOSFET upon the positive stress applied to the gate of the device, it is possible to notice that at a certain reading ($V_{G\text{-read}}$) fixed bias is possible to appreciate a decrease of the drain current. This delta on the drain current can be correlated to a variation of the threshold voltage from $V_{th}$ to $V_{th}$'. Hence, the $V_{th}$ variation can be argued with a single point measurement avoiding relaxing effects on the majority of the trapped charge.

By varying the gate bias stress, the proposed method allows to probe the whole semiconductor band gap and it differs from other methods that require dynamic adjustment of the measurements setup [12] and it is mainly based on simple geometric transformation to extract the $V_{th}$.

Fig. 3a shows the experimental procedure. The gate bias stress is ramped from $V_G$ = 0 V up to + 30 V then back toward the origin with steps of 5 V. At each step the $I_D$ is measured at $V_{G\text{-read}}$ = + 8V. In the second part of the experiment, the gate bias stress is ramped from down to - 25 V then back toward the origin. Each step of the cycling sequence is followed by a single point measure of the drain current as illustrated in Fig. 3b. The duration of the gate bias stress is varied and the results are depicted in Fig. 3c for stresses of 20, 200 e 2000 ms. As can be seen, the current values measured for negative gate bias stress saturated at $8\times10^{-8}$ A already at $V_G$ = -10 V for all the stress times. On the other hand, in the positive gate bias stress region it is possible to see that the drain current decreased increasing the gate bias stress. Furthermore, despite a negligible variation between the stress at 20 and 200 ms a significant variation it is observed for the 2000 ms (as highlighted by red dashed circle in Fig. 3c) stress suggesting the presence of slow states that need a long stress time to be perturbed. The $I_D$ variation can be converted in a $\Delta V_{th}$ as shown in Fig. 3d. It is possible to notice that the $\Delta V_{th}$ magnitude is up to 0.9 V. This result emphasizes the relevance of the measurement setup in order to collect the information coming from the wider possible spectrum of trapping states. Furthermore, closing the cycling stress loop toward the value of $V_G = 0$ V, it can be seen that the final $V_{th}$ value is different from the original one. This difference corresponds to a residual charge density trapped in the $SiO_2$/4H-SiC system, NIOTs stimulated during the stress procedure.

In order to get further insights on the $SiO_2$/4H-SiC interface physics, temperature dependent measurements elucidate among the thermally activated trapping mechanisms and the tunneling phenomena (that are nearly temperature independent). The $\Delta V_{th}$ (and consequently the amount of trapped charge) are evaluated after the gate bias stress at different temperatures up to 200 °C where each stress has 2000 ms of duration (Fig.4). For the interpretation of the experimental data it is important to comment some magnified details in Fig. 3d. At high positive bias, the p-type semiconductor is in inversion and the Fermi level crossed the bottom of the 4H-SiC conduction band (Fig. 5a). On the other hand, for large negative gate bias stress the $I_D$ is increased due to an electron de-trapping at the interface states due to the crossing of the Fermi level with the top of 4H-SiC valence band as illustrated in Fig. 5b with the p-type MOS structure in accumulation. Hence, the gate bias variation from – 25 V up to +30V allows the movement of the Fermi level across the whole 4H-SiC band gap. Hence, the total $V_{th}$ variation can be correlated with the total amount of the interface states $N_{it}$ from the top of the valence band to the bottom of the conduction band (Fig. 5).

During the cyclic stress procedure, also NIOTs may be stimulated. The NIOTs that are in proximity with the $SiO_2$/4H-SiC interface may be emitted similarly to the interface states resulting undistinguishable. On the other hand, the NIOTs that are located far from the $SiO_2$/4H-SiC interface (up to ~1.3 nm [5]) are slow enough to be measured. In fact, in Fig. 3c it is possible to notice that there is a gap between the starting and the ending point of the circular stress procedure indicating the presence of a residual charge trapped in the NIOTs. Furthermore, as can be noticed in Fig. 4, the total $V_{th}$ variation occurring in the device decreases increasing the measurement temperature.

Fig. 6 shows the estimated amount of $N_{it}$ and NIOTs from the experimental data depicted in Fig. 4 accordingly with the description illustrated in Fig. 3d. In particular, after the extraction of each value of $\Delta V_{th}$ they are converted in terms of the trapped charge according with the following relation:

$$N_{Trap} = \frac{\Delta V_{th} \kappa \varepsilon_0}{q t_{ox}} \tag{3}$$

where $q$ is the electron charge, $t_{ox}$ and $\kappa$ are the $SiO_2$ thickness and relative permittivity respectively and $\varepsilon_0$ is the vacuum permittivity. Finally, $N_{it}$ is defined as the $\Delta V_{th}$ that occurred from gate bias values of – 25 V and + 30 V, while NIOTs is defined as the $\Delta V_{th}$ that occurred at $V_G = 0$ V at the end of the cyclic stress (see Fig. 3d). As can be seen, the $N_{it}$ decreases increasing the temperature (from $6 \times 10^{11}$ cm$^{-2}$) while the NIOTs are nearly constant in the investigated temperature range (about $1 \times 10^{11}$ cm$^{-2}$). The nearly temperature independent NIOTs confirms that charging mechanism is likely due to a tunneling from the semiconductor into the oxide traps and vice versa [5]. On the other hand, the temperature behavior of the $N_{it}$ can be understood considering a different behavior of the interface states at different temperatures. In particular, considering that during the procedure the Fermi level is moved from the top of the valence band to the top of the conduction band, all the $N_{it}$ in the band gap are stimulated. Hence, the reduction of the amount of detected charge increasing the temperature can be interpreted considering the different retention capability of the traps at different temperatures under two possible mechanisms: thermal emission and/or trap-assisted tunneling. In other words, increasing the temperature the charges can be easily de-trapped from the charged states. In the insert of Fig. 6 shows the Arrhenius plot of the $N_{it}$ as a function of the reverse of the absolute temperature. The experimental data are in good agreement with a linear behavior having an activation energy $E_A$ of 0.1 eV. Since the most pronounced variations are observed at high positive gate bias stress, it is possible to argue that the larger amount of are located interface traps below the conduction band in good agreement with literature data [15,16,17,18]. In particular, it is possible to conclude that – accordingly with *Afanas'ev et al.* [16] – the $N_{it}$ with an activation energy of $E_A = 0.1$ eV can be associated to the intrinsic interface state defects of the $SiO_2$/4H-SiC system; i.e. oxygen related defects. On the other hand, the estimated NIOTs are most likely defects distributed in the insulator; i.e. residual carbon related defects [5].

In conclusion, a non-relaxing method to probe the threshold voltage instability in 4H-SiC MOSFETs has been used to separate the contribution of two different trapping mechanisms in the $SiO_2$/SiC system probing the whole 4H-SiC band gap. In particular, it has been found than one trapping mechanism is nearly temperature independent and it is most likely related to defect distributed in the insulating layer that are charged and discharged via tunneling. On the other hand, a thermally activated mechanism with an activation energy of 0.1 eV is found to be the most important trapping effect related to intrinsic defects of the $SiO_2$/4H-SiC system.

This work was carried out in the framework of the ECSEL JU project REACTION (first and euRopEAn siC eigTh Inches pilOt liNe), grant agreement no. 783158.

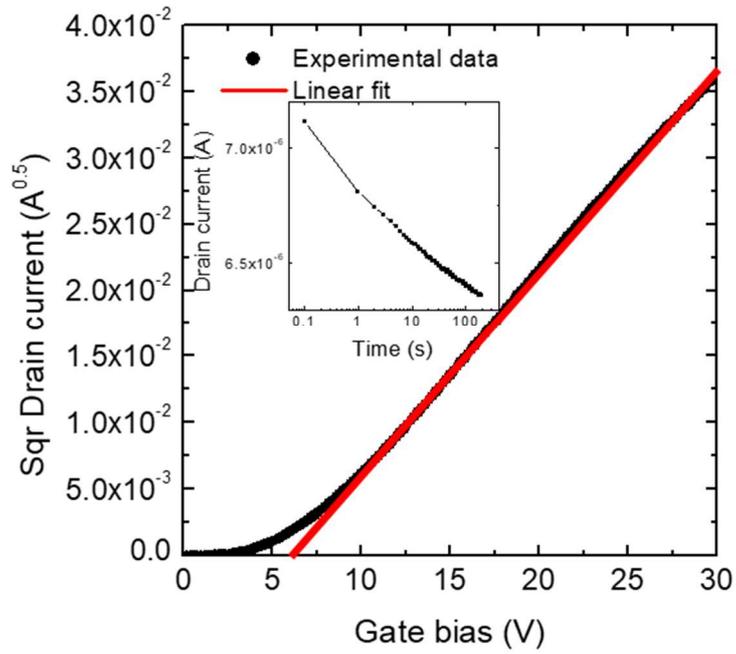

Fig. 1: Square root of the drain current as a function of the gate bias; the linear fit of the experimental data is used to determine the $V_{th}$. In the insert the drain current as a function of time at fixed value of the gate and drain terminals.

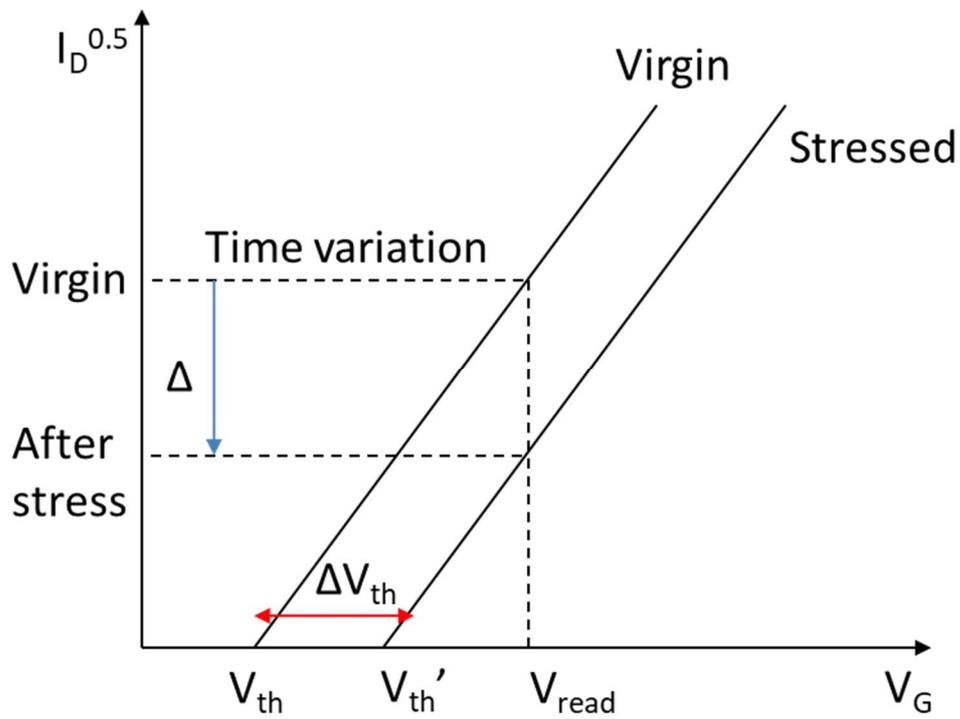

Fig. 2: Schematic illustration that explains how to extrapolate the $V_{th}$ variation from a single point measurement of the drain current at fixed gate bias.

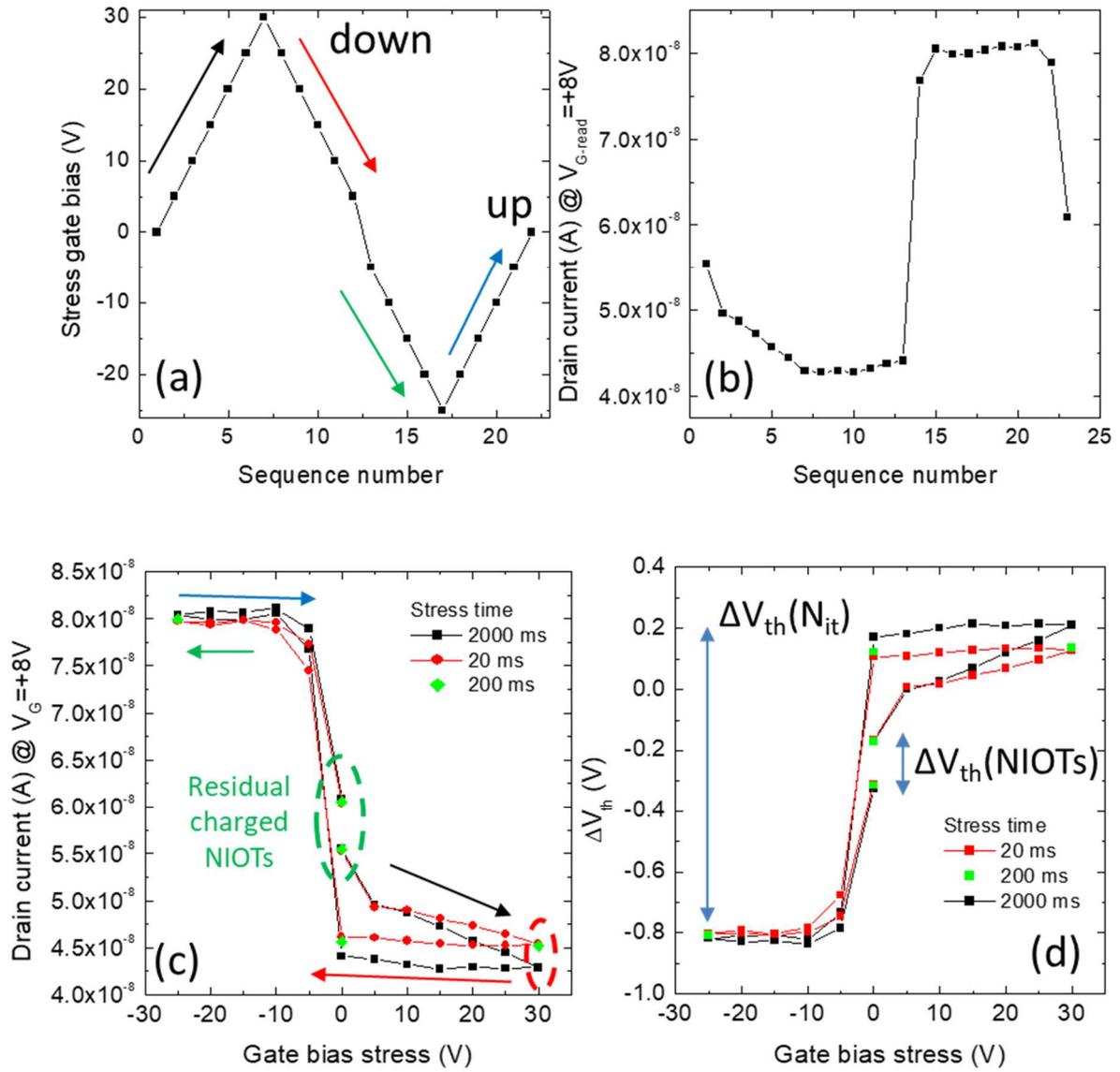

Fig. 3: (a) Schematic illustration of the cycling gate bias stress sequence procedure used to extrapolate the drain current variation (b) at a fixed read gate bias value. (c) Experimental data collected during the circular gate bias stress having duration of 20, 200 and 2000 ms. (d) Extrapolation of the $\Delta V_{th}$ measured and identification of the two trapping mechanisms.

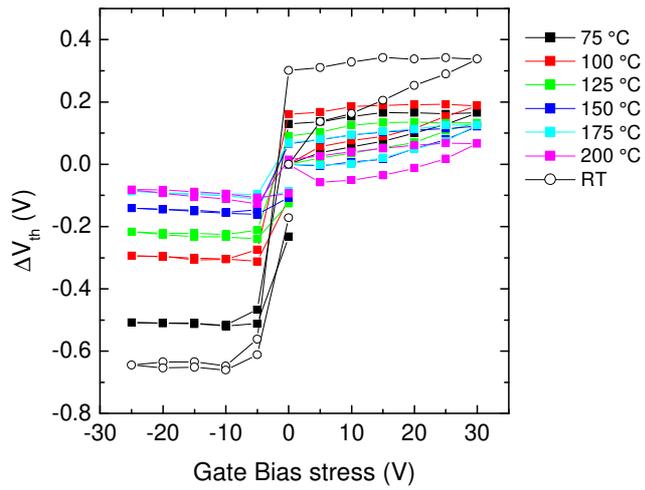

Fig. 4: (a) $I_D$-$V_G$ curves collected at different temperatures used as a reference to extrapolate the $\Delta V_{th}$ measured from the (b) circular gate bias stress performed at the same temperatures. (c) Extrapolation of the $\Delta V_{th}$ measured at different temperatures.

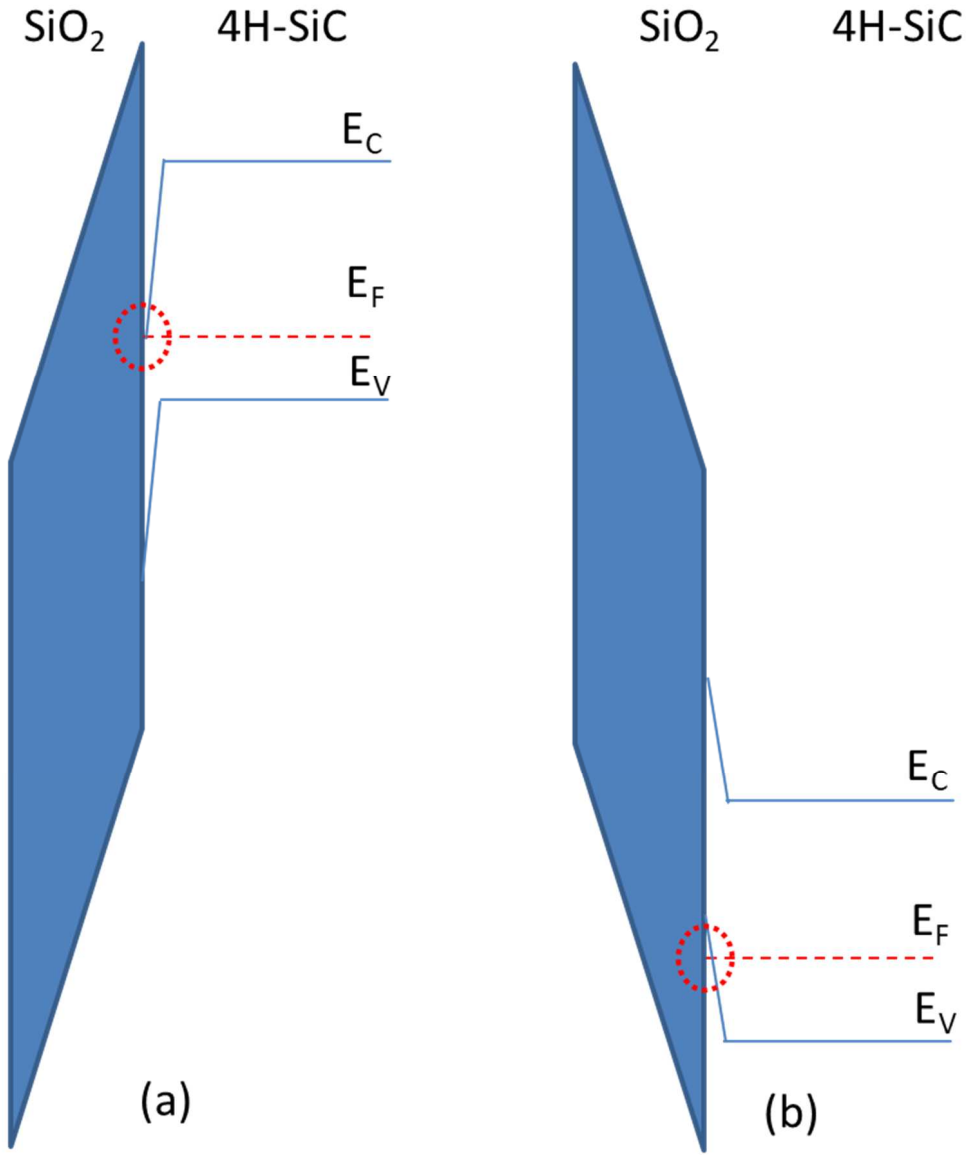

Fig. 5: (a) Schematic band diagram of the MOSFET in strong inversion with the Fermi level in contact with the bottom of the 4H-SiC conduction band. (b) Schematic band diagram of the MOSFET in strong accumulation with the Fermi level in contact with the top of the 4H-SiC valence band.

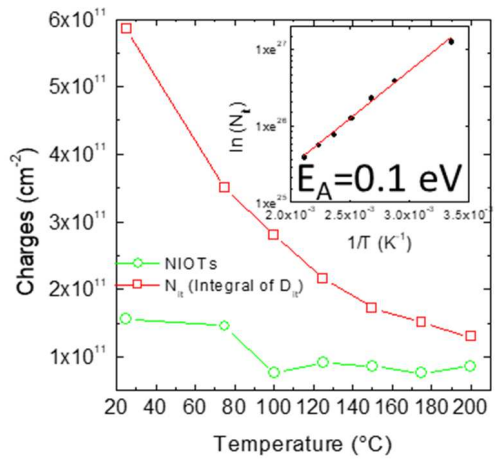

Fig. 6: Experimental value of the $N_{it}$ and NIOTs measured at different temperatures. Arrhenius plot of the $N_{it}$ in the insert.